\documentclass[showpacs,amsmath,
superscriptaddress,
twocolumn,
aps,prl]{revtex4}
\usepackage{graphicx}

\begin{document}

\title{Exciton diffusion in air-suspended single-walled carbon
nanotubes}
\author{S. Moritsubo}
\author{T. Murai}
\author{T. Shimada}
\affiliation{Institute of Engineering Innovation, 
The University of Tokyo, Tokyo 113-8656, Japan}
\author{Y. Murakami}
\affiliation{Global Edge Institute, Tokyo Institute of
Technology,
Tokyo 152-8550, Japan}
\author{S. Chiashi}
\author{S. Maruyama}
\affiliation{Department of Mechanical Engineering, 
The University of Tokyo, Tokyo 113-8656, Japan}
\author{Y. K. Kato}
\email[Corresponding author. ]{ykato@sogo.t.u-tokyo.ac.jp}
\affiliation{Institute of Engineering Innovation, 
The University of Tokyo, Tokyo 113-8656, Japan}
\affiliation{PRESTO, Japan Science and Technology Agency, 
Saitama 332-0012, Japan}

\begin{abstract}
Direct measurements of the diffusion length of excitons in air-suspended single-walled carbon nanotubes are reported. Photoluminescence microscopy is used to identify individual nanotubes and to determine their lengths and chiral indices. Exciton diffusion length is obtained by comparing the dependence of photoluminescence intensity on the nanotube length to numerical solutions of diffusion equations. We find that the diffusion length in these clean, as-grown nanotubes is significantly longer than those reported for micelle-encapsulated nanotubes. 
\end{abstract}
\maketitle
Optical properties of single-walled carbon nanotubes (SWCNTs)
are of importance because of their potential applications
in nanoscale photonics and optoelectronics \cite{Avouris:2008}, and exhibit
interesting physics that are unique to one-dimensional
systems. Limited screening of Coulomb interaction in SWCNTs
causes electron-hole pairs to form excitons with large binding
energies \cite{Wang:2005}, and these excitons play a central role in optical processes. 
There exists an
upper limit to the exciton density in SWCNTs \cite{Murakami:2009,Xiao:2010} caused by
exciton-exciton annihilation \cite{Wang:2004,Ma:2005, Matsuda:2008}. Since
the annihilation rate is determined by exciton diffusion 
\cite{Murakami:2009,Russo:2006,Luer:2009},
its elucidation is a key to understanding light emission processes and their efficiencies in SWCNTs.

Exciton diffusion is typically characterized by the diffusion length $L=\sqrt{D \tau}$ 
where $D$ is the diffusion coefficient and $\tau$ is the exciton lifetime.
Stepwise quenching of flourescence \cite{Cognet:2007} has yielded exciton excursion range
$\Lambda=2L=90$~nm, while $L=6$~nm has been reported from time-resolved measurements \cite{Luer:2009}. 
Recent near-field measurement has resulted in $\sqrt{2}L=100$~nm
\cite{Georgi:2009}. These measurements have been performed on micelle-encapsulated SWCNTs,
but it is expected that the transport properties of excitons are extremely sensitive 
to their surrounding environment.
The exciton diffusion length in clean, pristine SWCNTs has the potential to be considerably longer,
but the measurements done on suspended $(7, 5)$ nanotubes turned out to show $L_d=\sqrt{2}L=200$~nm \cite{Yoshikawa:2010}.

Here we report direct measurements of the diffusion length of
excitons in air-suspended SWCNTs. Individual nanotubes are identified by photoluminescence (PL)
imaging, while their lengths and chiral indices are determined 
by excitation spectroscopy and polarization measurements. With data obtained 
from 35 individual $(9,8)$ SWCNTs, we are able to extract the exciton
diffusion length $L$ by comparing the dependence of PL intensity
on the nanotube length with numerical solutions of diffusion
equations. We find that the diffusion length is at least 610~nm, which is
substantially longer than those reported for
micelle-encapsulated SWCNTs \cite{Cognet:2007,Luer:2009,Georgi:2009}. 
The apparent diffusion
length becomes shorter with higher excitation powers, consistent
with exciton-exciton annihilation effects.

\begin{figure}\includegraphics{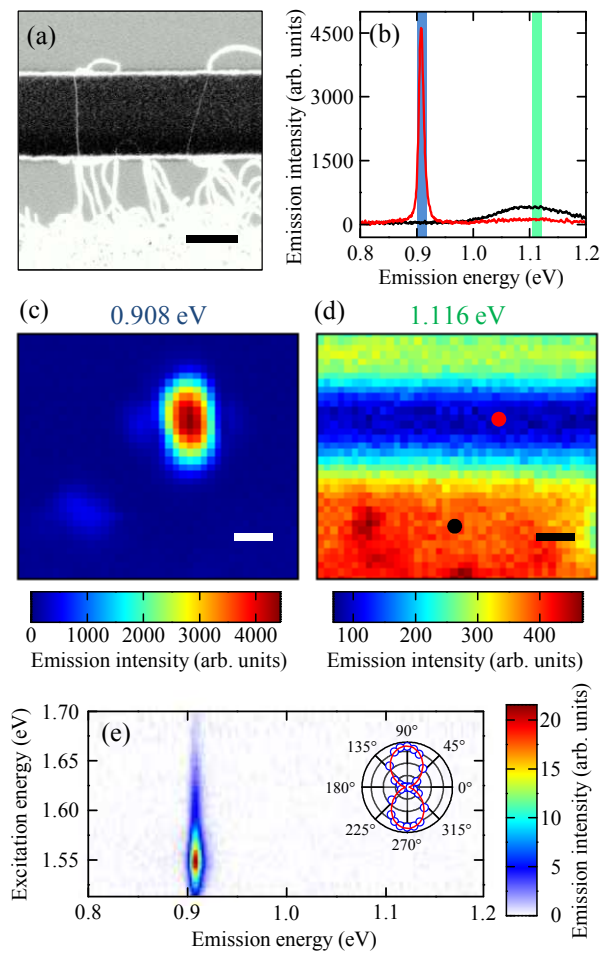}\caption{\label{fig1}
(a) A scanning electron microscope image of a typical sample.
The scale bar is 1~$\mu$m. 
(b) PL spectra for a
carbon nanotube (red) and Si substrate (black). Blue and green
shades indicate the 4~meV wide integration windows that are
used to obtain PL images at emission energies
of (c) 0.908~eV and (d) 1.116~eV, respectively. 
The red and black dots indicate the positions of the laser spot
where the red and black curves in (b) were taken, respectively.
The scale bars in (c) and (d) are 1~$\mu$m. 
For (b--d), excitation energy of 1.653~eV and excitation power
of 0.36 mW are used. (e) PL excitation
map for the same nanotube. Excitation power of 1.5 to 2.5~$\mu$W
is used. The spectra are corrected for the changes in excitation
power with tuning of the excitation energy. Inset is the laser
polarization angle dependence of the PL intensity
showing polarization $p=0.90$ for this nanotube. Blue circles are data and the red line is a fit. Taken at an
excitation energy of 1.653~eV and an excitation power of
2.24~$\mu$W.
}\end{figure}

In order to obtain suspended SWCNTs of various lengths, trenches
with widths ranging from 0.4~$\mu$m to 2.0~$\mu$m are prepared on (001)
SiO$_2$/Si substrates. Electron beam lithography and dry etching
processes are used to form the trenches, and an additional
electron beam lithography step is performed to define catalyst
areas next to the trenches. Silica supported Co/Mo catalyst
suspended in ethanol is spin-coated and lifted off. SWCNTs are
directly grown on these substrates by alcohol catalytic chemical
vapor deposition \cite{Maruyama:2002}. A scanning electron microscope image of a
typical sample is shown in Fig.~\ref{fig1}(a). We note that these
suspended as-grown SWCNTs are very clean, and exhibit excellent
optical and electrical properties \cite{Lefebvre:2003,Mann:2007,Cao:2005}.

A home-built laser-scanning confocal microscope system is used
for the PL measurements. In order to excite SWCNTs, an output of
a wavelength-tunable continuous-wave Ti:sapphire laser is
focused on the sample with a microscope objective lens. PL is
collected through a confocal pinhole corresponding to an
aperture with 3~$\mu$m diameter at the sample image plane. A fast
steering mirror allows lateral scanning of the laser spot for
acquiring PL images, and the laser polarization can be rotated
with a half-wave plate. PL spectra are collected with a single
grating spectrometer and a liquid nitrogen cooled InGaAs
photodiode array. All measurements are performed at room
temperature in air.

Suspended SWCNTs are identified by taking spectrally resolved PL
images. We raster the laser spot across the scan area on the
sample and take a PL spectrum at each position. By extracting the
emission intensity at the desired emission energy from these PL
spectra and replotting it in real space, PL images can be
constructed at any spectral position. Typical PL spectra from an
individual SWCNT and the substrate are shown in Fig.~\ref{fig1}(b). 
The bright sharp emission line near 0.9~eV is attributed to PL from
a suspended SWCNT, while PL from the Si substrate shows a broad
peak around 1.1~eV. PL images at emission energies corresponding
to the SWCNT [Fig.~\ref{fig1}(c)] and Si substrate [Fig.~\ref{fig1}(d)] unambiguously
show localized SWCNT emission at the trench position. If the
nanotube PL position does not coincide with the underlying
trench, we exclude those nanotubes from further measurements as
they may not be fully suspended. We also exclude nanotubes if
they show considerably lower emission intensity as those are
likely to have defects or surface contamination.

Once we find a suspended SWCNT with bright emission, we perform
PL excitation spectroscopy for chirality assignment. As shown in
Fig.~\ref{fig1}(e), only a single peak is visible throughout the
measurement range of excitation and emission energies, and we
determine the chirality of this nanotube to be (9,8) from
tabulated data \cite{Lefebvre:2004apa,Ohno:2006}. If we find two or more peaks in the PL
excitation spectra, those nanotubes are rejected since they may
be bundled.

Finally, PL intensity is measured as a function of polarization
angle of the excitation laser [Fig.~\ref{fig1}(e) inset]. We fit the data
to $I_0+I_1 \sin^2(\varphi+\varphi_0)$ where $I_0$ and $I_1$
are unpolarized and polarized PL intensity, respectively, $\varphi$ is the excitation polarization
angle, and $\varphi_0$ is the angle offset. From the fit parameters, we
compute the polarization $p=I_1/(I_0+I_1)$ and use it as a measure of the
straightness of the nanotube. Since uncertainties in the
nanotube length caused by bending is undesirable,
we limit ourselves to nanotubes with $p>0.5$ for the measurement
of the exciton diffusion length. Under the assumption that the
nanotube is relatively straight, the length $l$ of the suspended
portion of the nanotube is given by $l=w/\sin\varphi_0$ where $w$ is the
width of the trench.

\begin{figure}\includegraphics{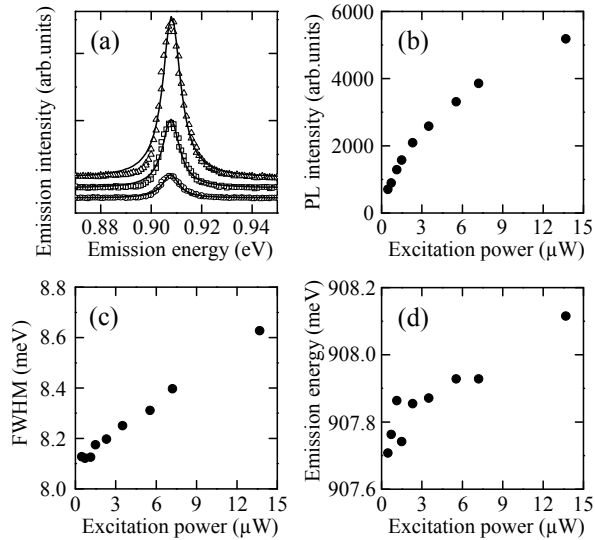}\caption{\label{fig2}
(a) Nanotube emission spectra at excitation powers of
0.47~$\mu$W (circles), 2.32~$\mu$W (squares), and 13.7~$\mu$W
(triangles). Data are taken at an excitation energy of 1.553~eV. 
Data are offset for clarity and lines are Lorentzian fits to data. The same
nanotube as in Fig.~\ref{fig1}(b--e) is used. (b), (c), and (d)
show photoluminescence intensity, full-width at half-maximum
(FWHM) of the emission line, and emission energy as a function
of excitation power, respectively.
}\end{figure}

Following such careful selection and characterization
procedures, we have investigated 35 individual SWCNTs with a
chiral index of (9,8). We focus on a single chirality in order
to avoid any chirality dependent effects. For each of these
nanotubes, we have collected a series of PL spectra as a function of
excitation power $P$. These measurements are done with the laser
spot at the center of the nanotube, 
the laser polarization adjusted for maximum emission intensity, and the excitation energy
tuned to the peak of the resonance. Typical data are shown in Fig.~\ref{fig2}(a), 
and we fit the nanotube peak with a Lorentzian function in
order to extract the peak height, width, and position. We
calculate the peak area from the fit parameters, and use this as
the measure of the PL intensity. It shows a sublinear behavior
[Fig.~\ref{fig2}(b)], as expected from exciton-exciton annihilation
\cite{Murakami:2009,Xiao:2010,Matsuda:2008}. 
We can also estimate
the amount of laser induced heating from the broadening of the linewidth [Fig.~\ref{fig2}(c)] 
by comparing with previous temperature dependent measurements 
\cite{Matsuda:2008,Yoshikawa:2009,Lefebvre:2004prb}. Heating may
also influence the diffusion constant \cite{Yoshikawa:2010}, 
but we find that the increase in the temperature is
less than 30~K in all of the nanotubes under investigation.
We also observe the blueshift of the emission line [Fig.~\ref{fig2}(d)], 
which may be related to gas adsorption \cite{Finnie:2005,Chiashi:2008}.

Since we want to analyze the dependence of the PL intensity on
$l$ to obtain the diffusion length, we have
simulated the exciton density profile based on a 
steady-state one-dimensional diffusion equation given by 
\[
D \frac{d^2 n(z)}{dz^2}-\frac{n(z)}{\tau} + \Gamma(z)=0,
\]
where 
$n(z)$ is the exciton density, $z$ is the position on the nanotube,
 and $\Gamma(z)$ is the exciton
generation rate. This equation does not explicitly contain the exciton-exciton annihilation term, 
but to first order approximation, such an effect can be described by a shorter $\tau$ within this simple model.
We set the origin to be at the center of the
nanotube and impose boundary conditions to be $n(\pm l/2)=0$,
assuming that the quenching of PL due to interaction with the
substrate is sufficiently strong. Since the exciton generation rate
is proportional to the laser intensity profile, we let $\Gamma(z)=
\Gamma_0 \exp(-z^2/\sigma^2)$, where $\Gamma_0$ is a proportionality
constant and $\sigma=520$~nm is the width of the laser spot. The
diffusion equation becomes 
\[
L^2 \frac{d^2 n(z)}{dz^2}-n(z)+N \exp(-\frac{z^2}{\sigma^2})=0,
\]
where $N =  \Gamma_0 \tau$ is a constant.

\begin{figure}\includegraphics{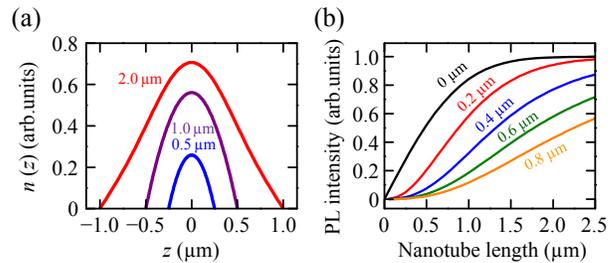}\caption{\label{fig3}
(a) Simulated exciton density spatial profile for 
$l=0.5$~$\mu$m (blue), 1.0~$\mu$m (purple), and
2.0~$\mu$m (red). Exciton diffusion length $L=0.3$~$\mu$m is
used for obtaining these curves. (b) Simulated PL
intensity as a function of $l$ for 
$L= 0.0$~$\mu$m (black), 0.2~$\mu$m (red), 0.4~$\mu$m (blue),
0.6~$\mu$m (green), and 0.8~$\mu$m (orange).
}\end{figure}

We numerically solve this equation to obtain the exciton density
profile, and the results for $L=0.3$~$\mu$m are plotted in Fig.~\ref{fig3}(a).
For $l$ shorter or comparable to $L$, the exciton density profile is nearly parabolic, indicating
that majority of excitons diffuse to the unsuspended regions
before recombining. Since the nonradiative recombination within
unsuspended region efficiently removes excitons, the density of
excitons stays low compared to longer nanotubes. As the nanotube
length gets longer, the exciton density increases until it
saturates when the nanotube length becomes long enough compared to $2(\sigma + L)$.
In such a situation, most of the excitons recombine before they diffuse
out to the unsuspended part, so that $l$
does not play a role.

In order to compute how the PL intensity changes with $l$, 
we integrate the exciton density profile to obtain the
total number of excitons. We simulate the PL intensity for a
range of nanotube length and a series of diffusion lengths, and
plot them in Fig.~\ref{fig3}(b). If the diffusion length is very short, the saturation
of the PL intensity is expected for nanotube length longer than
the laser spot size. As the diffusion length gets longer, the
transition to the linear behavior shifts to longer nanotube
length, and saturation would not be observed.

\begin{figure}\includegraphics{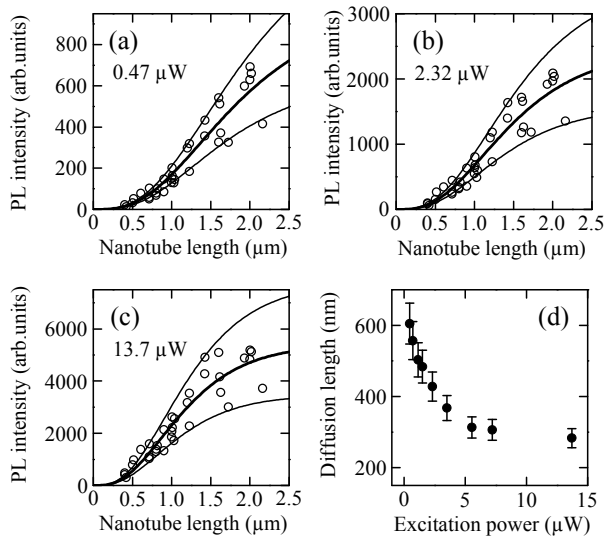}\caption{\label{fig4}
(a), (b), and (c) show measured photoluminescence intensity as a
function of nanotube length for $P=0.47$~$\mu$W, 2.32~$\mu$W, and 13.7~$\mu$W, respectively. 
Polarization dependence of the detection efficiency has been corrected.
The thick solid lines represent the best fits and the thin lines
correspond to simulation results with exciton lifetime scaled
by $\pm 20$~\% while all the other parameters were fixed. 
The best fits give $L=610$~nm, 430~nm, and 280~nm for (a), (b), and(c), respectively.
(d) Excitation power dependence of exciton diffusion length.
}\end{figure}

Now we compare the $l$ dependence of the measured PL
intensity to the simulation [Fig.~\ref{fig4}(a--c)]. We perform least
square fits to the experimental data by looking for optimum
values for $L$ and $N$. At all
excitation powers, we find reasonable agreement between the data
and the simulation.
Although the experimental data show some dispersion from the fit, it
is expected that there are some tube-to-tube variations in the
PL intensity. Such inhomogeneities have been observed in PL
imaging of very long SWCNTs \cite{Lefebvre:2006}, and can result from changes in
the exciton lifetime $\tau$ induced by gas adsorption, contamination, 
or defects. We have computed such an effect, and found that our data falls into the band between the
simulation results with lifetime set to $\pm 20$~\% of the best fit. Taking such uncertainties
in $\tau$ as the error in the determination of $L$, 
we plot the dependence of the diffusion length as a
function of $P$ in Fig.~\ref{fig4}(d).

The apparent diffusion length decreases with increasing $P$,
which can be qualitatively explained by a reduction of $\tau$ caused by
exciton-exciton annihilation. However, the $z$ and $l$ dependences of $\tau$ is
not accounted for in our simple model, so it may not be accurate at high
excitation powers. A more rigorous modeling is required to clarify the effects
of exciton-exciton annihilation on $L$. Nevertheless, such effects should 
be small at lower powers. We find $L=610$~nm for the lowest $P$, 
and an extrapolation of the data down to $P=0$~$\mu$W
suggests even longer $L$.
The observed diffusion length is much longer compared to micelle-encapsulated SWCNTs \cite{Luer:2009, Cognet:2007, Georgi:2009}. 
It is possible that surfactants cause additional
exciton scattering, and sonication may introduce defects in
those samples.

The diffusion length can give us some insight to transport
properties of excitons. From $L= 610$~nm, we
obtain $D=44$~cm$^2$~s$^{-1}$ for $\tau=85$~ps \cite{Xiao:2010}.
Using the Einstein relation $\mu kT = eD$, where $\mu$ is the exciton mobility, 
$k$ is the Boltzmann constant, and $e$ is the electronic charge, we find 
$\mu=1.7 \times 10^3$~cm$^2$~V$^{-1}$~s$^{-1}$. This is comparable to reported values of
carrier mobilities in SWCNTs \cite{Javey:2002,Durkop:2004,Zhou:2005}, although different scattering mechanisms and effective masses should be considered in general. 

In summary, the diffusion length of excitons in air-suspended $(9,8)$ SWCNTs
has been measured to be as long as 610~nm. At higher
excitation powers, the apparent diffusion length becomes shorter due to
exciton-exciton annihilation. Long diffusion lengths are
favorable for fabricating single photon sources from SWCNTs
\cite{Hogele:2008}, because such a device will need to have a length less than
the exciton diffusion length to ensure annihilation of excess
excitons.

\begin{acknowledgments}
We acknowledge support from the Iketani Science and Technology
Foundation, Murata Science Foundation, Research Foundation for
Opto-Science and Technology, Center for Nano Lithography \&
Analysis at The University of Tokyo, and Photon Frontier
Network Program of MEXT, Japan.
\end{acknowledgments}

\end{document}